\documentclass[nofootinbib,twocolumn,showpacs,preprintnumbers,pre,aps,superscriptaddress]{revtex4-1}

\usepackage[dvipdfmx]{graphicx}
\usepackage{bm}
\usepackage{amsmath}
\usepackage{amssymb}
\usepackage{color}

\begin{document}

\title{The Onsager-Machlup Integral for Non-reciprocal Systems with Odd Elasticity}%

\author{Kento Yasuda}
\affiliation{
Research Institute for Mathematical Sciences, Kyoto University, Kyoto 606-8502, Japan}

\author{Akira Kobayashi}
\author{Li-Shing Lin}
\author{Yuto Hosaka}
\author{Isamu Sou}
\affiliation{
Department of Chemistry, Graduate School of Science, Tokyo Metropolitan University, Tokyo 192-0397, Japan}

\author{Shigeyuki Komura}\email{komura@wiucas.ac.cn}
\affiliation{
Department of Chemistry, Graduate School of Science, Tokyo Metropolitan University, Tokyo 192-0397, Japan}
\affiliation{
Wenzhou Institute, University of Chinese Academy of Sciences, Wenzhou, Zhejiang 325001, China}
\affiliation{
Oujiang Laboratory, Wenzhou, Zhejiang 325000, China}

\date{\today}

\begin{abstract}
The variational principle of the Onsager-Machlup integral is used to describe the stochastic 
dynamics of a micromachine, such as an enzyme, characterized by odd elasticity. 
The obtained most probable path is found to become non-reciprocal in the presence 
of odd elasticity and is further related to the entropy production.
\end{abstract}

\maketitle

\begin{figure*}[tbh]
\begin{center}
\includegraphics[scale=0.26]{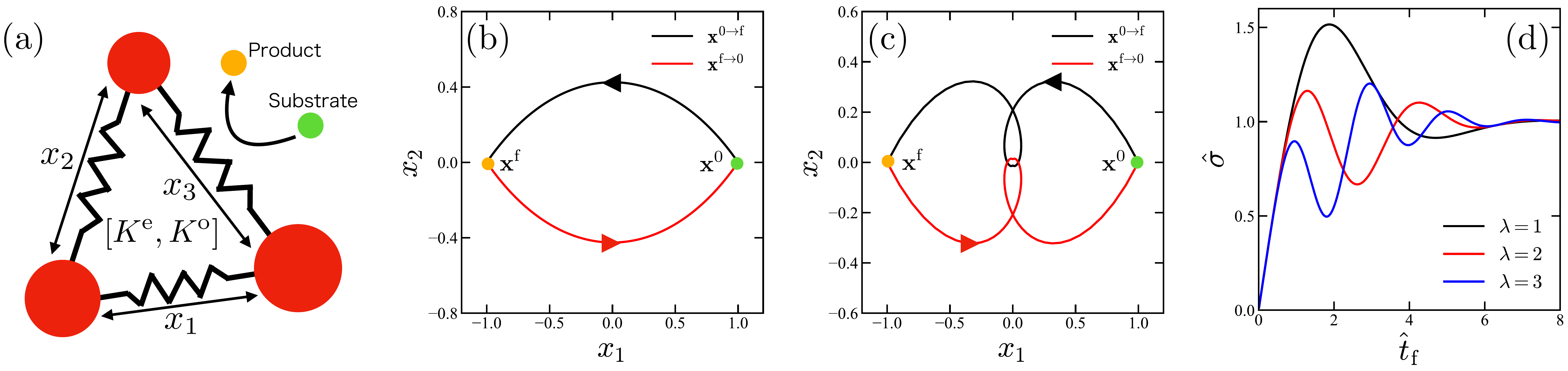}
\end{center}
\caption{
(Color online)
(a) A coarse-grained model of an enzyme consisting of domains that are connected to springs.
A substrate (green circle) changes into a product (orange circle) via a catalytic chemical reaction.
The distances between the domains are represented by $x_i(t)$.
Each spring is characterized by the even elastic constant tensor $K_{ij}^{\rm e}$ and the odd 
elastic constant tensor $K_{ij}^{\rm o}$.
(b), (c) The most probable outward path $\mathbf x^{0 \rightarrow {\rm f}}$ (black line) 
and the return path $\mathbf x^{{\rm f} \rightarrow 0}$ (red line) when $N=2$ and 
$\lambda=1$.
The initial and final conditions are $\mathbf x^0 = (1,0)$ (green circles) and 
$\mathbf x^{\rm f} = (-1,0)$ (orange circles), respectively, whereas the final times are 
$\hat t_{\rm f}=1$ in (b) and $\hat t_{\rm f}=10$ in (c).
(d) The dimensionless entropy production $\hat \sigma$ corresponding to the most probable 
non-reciprocal cyclic path as a function of the final time $\hat t_{\rm f}$ for $\lambda=1, 2, 3$.
}
\label{model}
\end{figure*}

Structural changes in enzymes or proteins are essential for biological functions 
and have attracted extensive attention~\cite{Yasuda21}.
Such dynamical transitions are associated with various active processes including 
catalytic chemical reactions.
Because thermal fluctuations act as a driving force, structural changes can be 
regarded as rare events that occur stochastically. 
To describe the dynamics of rare events, several concepts such as
the path probability and the Onsager-Machlup (OM) integral were employed~\cite{Onsager53}.
Recently, Scheibner et al.\ introduced the concept of odd elasticity to describe 
non-conserved interactions in active materials~\cite{Scheibner20}.
They showed that the active moduli quantify the amount of work extracted during quasistatic 
strain cycles.

In the present work, we investigate how the presence of odd elasticity in a micromachine 
influences the most probable path derived by the variation of the OM integral, namely,   
the OM variational principle~\cite{Doi19}. 
The most probable outward path is shown to differ from the most probable return path in 
active micromachines; hence, the entire process becomes non-reciprocal owing to odd 
elasticity.
Using the fluctuation theorem~\cite{Seifert12}, we also estimate the entropy production which 
is generated by the most probable cyclic path.
Our study is useful for understanding the dynamics of the binding and dissociation processes 
between the enzyme and substrate molecules.
Furthermore, this research suggests that non-equilibrium processes, including catalytic chemical 
reactions, can generally be described by the concept of odd elasticity in active systems.

Let us introduce the $N$-dimensional time-dependent state vector $\mathbf x(t)$
whose components are $x_i(t)$ ($i=1,2,\cdots,N$).
These variables describe, for example, the distances between the domains in an enzyme
as shown in Fig.~\ref{model}(a).
We assume that $x_i(t)$ obeys the following linear Langevin equation~\cite{Yasuda21b}
\begin{align}
\dot x_i(t)=-\mu_{ij}K_{jk}x_k(t)+F_{ij}\xi_j(t), 
\label{OverLanEq}
\end{align}
where $\dot x_i=d x_i/dt$.
The transport coefficient tensor $\mu_{ij}$ is symmetric ($\mu_{ij}=\mu_{ji}$) and positive 
definite owing to Onsager's reciprocal relations and the second law of thermodynamics, 
respectively~\cite{DoiBook}.

In the above equation, $K_{ij}$ is the elastic constant tensor.
For ordinary passive situations, $K_{ij}$ should be symmetric because the elastic forces are conservative.
However, for active systems with non-conservative interactions, $K_{ij}$ can have an anti-symmetric 
part that corresponds to odd elasticity~\cite{Scheibner20,Yasuda21b}.
Hence $K_{ij}$ can generally be written as 
\begin{align}
K_{ij}=K_{ij}^{\rm e}+K_{ij}^{\rm o},
\label{elasticity}
\end{align}
where the symmetric (even) part and the anti-symmetric (odd) part satisfy $K_{ij}^{\rm e}=K_{ji}^{\rm e}$ 
and  $K_{ij}^{\rm o}=-K_{ji}^{\rm o}$, respectively.
Within a coarse-grained description, non-equilibrium processes in active systems can generally be 
described by odd elasticity.
For example, conformational changes of an enzyme are driven by catalytic chemical reactions, leading 
to non-conserved interactions between domains. 
The outcome of such non-equilibrium interactions can be considered by the odd part of 
the elastic constant tensor $K_{ij}^{\rm o}$~\cite{Yasuda21b}.

Moreover, $\xi_i$ in Eq.~(\ref{OverLanEq}) represents $N$-dimensional Gaussian white noise with a zero mean
$\langle \xi_i(t) \rangle=0$, and its correlations satisfy the relation
$\langle \xi_i(t) \xi_j(t')\rangle=\delta_{ij}\delta(t-t')$. 
The tensor $F_{ij}$ represents the noise strength and is further related to the diffusion 
tensor by $D_{ij}=F_{ik}F_{kj}/2$.
In this study, stochastic transition processes driven by thermal fluctuations are considered 
and the relation $D_{ij}=k_{\rm B}T\mu_{ij}$ is assumed, where $k_{\rm B}$ is 
the Boltzmann constant and $T$ is the temperature.
This means that the random force is determined by the thermal motion of the surrounding fluid 
molecules and will not be affected by the force acting on the domains~\cite{DoiBook}.
In terms of the probability distribution function $\mathcal P(\mathbf x,t)$, the Fokker-Planck equation, 
equivalent to Eq.~(\ref{OverLanEq}), is written as~\cite{RiskenBook}  
\begin{align}
\dot {\mathcal P}(\mathbf x,t)=\mathcal L(\mathbf x,t) \mathcal P(\mathbf x,t),~
\mathcal L(\mathbf x,t)=\partial_i \mu_{ij}K_{jk}x_k +D_{ij}\partial_i \partial_j,  
\label{FPeq}
\end{align}
where $\mathcal L$ is the Fokker-Planck operator and $\partial_i=\partial/\partial x_i$.

Next, we discuss the transition dynamics that occur, for example, in an enzyme as shown in 
Fig.~\ref{model}(a).
When a catalytic chemical reaction takes place, the dissociation process between the enzyme 
and the substrate molecules follows a configurational path that differs from the binding 
process between them.
To describe such transition dynamics, we consider the path probability, namely, the probability 
of a specific stochastic trajectory~\cite{Onsager53}.

When the initial condition is $x_i=x_i^0$ at $t=0$, the path probability 
$P[\mathbf x(t)|\mathbf x^0]$ during the time interval $0 \le t \le t_{\rm f}$, where $t_{\rm f}$ 
is the final time, is obtained from the product of the conditional probability distribution functions 
$\mathcal P(\mathbf x,t|\mathbf x',t')$ for a small time separation $t-t'$.
With the use of Eq.~(\ref{FPeq}), it is known that the path probability is expressed as~\cite{RiskenBook} 
\begin{align}
P[\mathbf x(t)|\mathbf x^0]=\mathcal N\exp\left(-\frac{\mathcal O[\mathbf x(t)]}{2k_{\rm B}T}\right),
\label{PathProb}
\end{align}
where $\mathcal N$ is the normalization constant fixed by the condition
$\int \mathcal D\mathbf x\, P[\mathbf x(t)|\mathbf x^0]=1$ and $\int \mathcal D\mathbf x$
indicates integration over all paths.
In Eq.~(\ref{PathProb}), $\mathcal O[\mathbf x(t)]$ is the OM integral defined by~\cite{Onsager53}
\begin{align}
\mathcal O[\mathbf x(t)]&=\frac{k_{\rm B}T}{2}\int_{0}^{t_{\rm f}}dt\,D_{ij}^{-1}[\dot x_i+\mu_{ik}K_{kl}x_l]\nonumber\\
&\times[\dot x_j+\mu_{jm}K_{m n}x_n],
\label{OMint}
\end{align}
where $D_{ij}^{-1}$ is the inverse matrix of $D_{ij}$.

As clearly seen in Eq.~(\ref{PathProb}), a transition path that minimizes the OM integral is realized 
with the highest probability.
In other words, the most probable transition path can be obtained by requiring the first variation of 
the OM integral to vanish, i.e., $\delta \mathcal O[\mathbf x(t)]=0$~\cite{Doi19}.
Taking the variation of Eq.~(\ref{OMint}) with respect to $x_i(t)$ yields the following differential equation 
for the most probable transition path:
\begin{align}
\ddot x_i(t)+\mu_{ij}(K_{jk}-K_{kj})\dot x_k(t)-\mu_{ij}K_{kj}\mu_{kl} K_{lm}x_m(t)=0.
\label{MPPatheq}
\end{align}

Given the initial ($x_i^0$ at $t=0$) and the final ($x_i^{\rm f}$ at $t=t_{\rm f}$) conditions, the above 
equation can be solved for $x_i(t)$. 
Importantly, the coefficient of $\dot x_k(t)$ in Eq.~(\ref{MPPatheq}) is 
proportional to the odd elastic constant $K_{ij}^{\rm o}=(K_{ij}-K_{ji})/2$.
In other words, the presence of odd elasticity breaks the time-reversal symmetry of the equation.
Consequently, the outward and return processes of the most probable path generally 
differ, leading to a non-reciprocal trajectory in state space. 
This is the main message of this study.

As a simple example, the most probable path of an active system with only two degrees 
of freedom ($N=2$) is discussed.
Additionally, we assume that the transport coefficient tensor and elastic constant tensor are given by 
$\mu_{ij}=\mu\delta_{ij}$, $K_{ij}^{\rm e}=K^{\rm e}\delta_{ij}$, and $K_{ij}^{\rm o}=K^{\rm o}\epsilon_{ij}$, 
where $\epsilon_{ij}$ is the 2D Levi-Civita tensor with $\epsilon_{11}=\epsilon_{22}=0$ and 
$\epsilon_{12}=-\epsilon_{21}=1$.
Then, the analytic solution of the most probable outward path 
($\mathbf x^0 \rightarrow \mathbf x^{\rm f}$) is obtained as 
\begin{align}
x_i^{0 \rightarrow {\rm f}}(\hat t)&=
-\frac{\sinh(\hat t-\hat t_{\rm f})}{\sinh(\hat t_{\rm f})}A_{ij}(\lambda\hat t)x_j^0\nonumber\\
&+\frac{\sinh(\hat t)}{\sinh(\hat t_{\rm f})}A_{ij}(\lambda(\hat t-\hat t_{\rm f}))x_j^{\rm f},
\label{MPPsol}
\end{align}
where $A_{ij}(t)=\delta_{ij} \cos t-\epsilon_{ij} \sin t$, 
and the dimensionless quantities 
$\hat t=t \mu K^{\rm e}$, $\hat t_{\rm f}=t_{\rm f} \mu K^{\rm e}$, 
and $\lambda=K^{\rm o}/K^{\rm e}$ are introduced.
Notice that $A_{ij}$ satisfies the relation $A_{ij}(t)=A_{ji}(-t)$.

Then, let us exchange the initial and the final conditions, and consider the most probable trajectory 
$x_i^{{\rm f} \rightarrow 0}(\hat t)$ for the return path
($\mathbf x^{\rm f} \rightarrow \mathbf x^0$).
When $\lambda=0$ and hence $A_{ij}=A_{ji}$, we have 
$x_i^{0 \rightarrow {\rm f}}(\hat t)=x_i^{{\rm f} \rightarrow 0}(\hat t_{\rm f}-\hat t)$, indicating 
that the entire combined process ($\mathbf x^0 \rightarrow \mathbf x^{\rm f} \rightarrow \mathbf x^0$) 
is reciprocal. 
When $\lambda\ne 0$ and hence $A_{ij}\ne A_{ji}$, on the other hand, the process becomes non-reciprocal.
In Figs.~\ref{model}(b) and (c), we plot on the $(x_1, x_2)$-plane the outward path 
$x_i^{0 \rightarrow {\rm f}}(\hat t)$ (black line) and the return path $x_i^{{\rm f} \rightarrow 0}(\hat t)$ 
(red line) in the presence of odd elasticity ($\lambda=1$).
It is explicitly revealed that these two paths do not coincide; hence, the entire trajectory is 
non-reciprocal when $\lambda \neq 0$.

The path probability $P[\mathbf x(t)|\mathbf x^0]$ is closely related to the entropy production $\sigma$
along the stochastic path $x_i(t)$ (not necessarily the most probable path).
According to the fluctuation theorem,   
$P[\mathbf x(t)|\mathbf x^0]/P[\mathbf x^\mathrm{rev}(t)|\mathbf x^{\rm f}]=\exp(\sigma/k_B)$,
where $x_i^\mathrm{rev}(t)=x_i(t_{\rm f}-t)$ is the time-reversed backward path~\cite{Seifert12,Mahault21}.
With the use of Eqs.~(\ref{PathProb}) and (\ref{OMint}) for general $N$, $\sigma$ can be obtained as 
\begin{align}
\sigma = - \frac{K_{ij}}{T}\int_0^{t_{\rm f}}dt\,\dot x_i(t) x_j(t).
\label{sigma1}
\end{align}
For a cyclic path, only the odd elastic constant gives rise to a non-conservative force, and 
Eq.~(\ref{sigma1}) further reduces to 
\begin{align}
\sigma = -\frac{K_{ij}^{\rm o}}{T} \oint dx_i\, x_j. 
\label{sigma2}
\end{align}
Note that the above line integral corresponds to the area enclosed by a closed path on the 
$(x_i, x_j)$-plane.

Finally, we calculate the entropy production for the most probable non-reciprocal cyclic paths as 
shown in Figs.~\ref{model}(b) and (c) when $N=2$.
In Fig.~\ref{model}(d), a plot of the dimensionless entropy production 
$\hat{\sigma}=\sigma T/(2K^{\rm e} \lambda^2)$ as a function of  
$\hat t_{\rm f}$ is shown for different values of $\lambda$. 
Although the dependence of $\hat{\sigma}$ on $\hat t_{\rm f}$ is highly non-monotonic, it is interesting 
to note that $\hat{\sigma}$ takes the maximum values, which also depend on the value of $\lambda$.

For stochastic systems, it was reported that probability flux in closed loops is possible 
in a non-equilibrium steady state~\cite{Battle16,Sou19}.
While the probability flux can predict only short-time dynamics, the most probable path
contains information on the long-time and global behavior of the system.

K.Y.\ thanks K.\ Ishimoto for useful discussions.
K.Y.\ and Y.H.\ acknowledge the support by a Grant-in-Aid for JSPS Fellows (Grant Nos.\ 21J00096
and 19J20271) from the JSPS.
S.K.\ acknowledges the support by a Grant-in-Aid for Scientific Research (C) (Grant No.\ 19K03765) 
from the JSPS, and support by a Grant-in-Aid for Scientific Research on Innovative Areas
``Information Physics of Living Matters'' (Grant No.\ 20H05538) from the MEXT of Japan.


\end{document}